\documentclass[11pt]{article}
\usepackage[english]{babel}
\usepackage{amsmath}
\usepackage{amsfonts}
\usepackage{multicol}
\usepackage{stmaryrd}
\font\tensym=msbm10
\font\sevensym=msbm7
\font\fivesym=msbm5

\newfam\symfam
\textfont\symfam=\tensym
\scriptfont\symfam=\sevensym
\scriptscriptfont\symfam=\fivesym
\def\sym{\fam\symfam\tensym}

\def\ss{\smallskip}

\def\mhs{\hbox to 1cm{}}

\def\hs{\hbox to 3mm{}}

\newtheorem{proposition}{Proposition}

\newtheorem{example}{Example}

\newtheorem{note}[example]{Note}

\def\mref#1{(\ref{#1})}
\def\Proof{\noindent {\bf Proof.} }

\def\pointir{\unskip . --- \ignorespaces}
\def\ra{\rightarrow}

\newcommand{\ostar}{\framebox(6.5,6.5){$*$}}

\def\al{\alpha}
\def\be{\beta}

\def\ga{\gamma}

\def\Si{\Sigma}

\def\T{{\cal T}}

\def\N{{\sym N}}

\def\R{{\sym R}}

\title{{\bf Tables, Memorized Semirings and Applications}}
\author{
Cyrille Bertelle
\footnote{LIH, Laboratoire d'Informatique du Havre, 
25 rue Philippe Lebon, 
BP 540 
76058 Le Havre cedex, France
},\\
G\'erard H. E. Duchamp
\footnote{LIPN, 
Institut Galil\'ee - Universit\'e Paris XIII
99, avenue Jean-Baptiste Cl\'ement,  
93430 Villetaneuse, France}\\ and\\ 
Khalaf Khatatneh\footnote{LIFAR, Facult\'e des Sciences et des Techniques,
76821 Mont-Saint-Aignan Cedex, France}
\thanks{\tt cyrille.bertelle@univ-lehavre.fr, gerard.duchamp@lipn.univ-paris13.fr, khalaf.khatatneh@univ-rouen.fr}
}

\date{}
\begin{document}
\maketitle

\begin{multicols}{2}

\ss
{\bf Keywords:} Tables, $k$-subsets, efficient data structures, efficient algebraic structures.

\section{Introduction}
The following is intended to be a contribution in the area of what could be called 
{\it efficient algebraic structures}  or {\it efficient data structures}.  In fact, we define 
and construct a new data structure, the tables, 
which are special kinds of two-raws arrays. 
The first raw is filled with words and the second with some coefficients. 
This structure generalizes the (finite) $k$-sets sets of Eilenberg \cite{Ei}, it is versatile 
(one can vary the letters, the words and the coefficients), 
easily implemented and fast computable. Varying the scalars and the operations on them, 
one can obtain many different structures and, among them, semirings. 
Examples will be provided and worked out in full detail.\\
Here, we present a new semiring (with several semiring structures) which can 
be applied to the needs of automatic processing multi-agents behaviour problems. 
The purpose of this account/paper is to present also the basic elements of this new structures 
from a combinatorial point of view. These structures present a bunch of properties. They will be 
endowed with several laws namely : Sum, Hadamard product, Cauchy product, Fuzzy operations 
(min, max, complemented product)
Two groups of applications are presented.\\
The first group is linked to the process of ``forgetting'' information in the tables and then obtaining, 
for instance, a memorized semiring.The latter is specially suited to solve the {\it shortest path with addresses} 
problem by repeated squaring over matrices with entries in this semiring.\\
The second, linked to multi-agent systems, is announced by showing a methodology to manage emergent 
organization from individual behaviour models.

\section{Description of the data structure}

\subsection{Tables and operations on tables}\label{table}

The input alphabet being set by the automaton under consideration, we will here rather focus on the definition of semirings providing 
transition coefficients. For convenience, we first begin with various laws on $\R_+:=[0,+\infty[$ including 
\begin{enumerate}
\item $+$ (ordinary sum)
\item $\times$ (ordinary product)
\item min (if over $[0,1]$, with neutral $1$, otherwise must be extended to $[0,+\infty]$ and then, with neutral $+\infty$) or max
\item $+_a$ defined by $x+_a y:=log_a(a^x+a^y)$ ($a>0$)
\item $+_{[n]}$ (H\"older laws) defined by $x+_{[n]} y:=\sqrt[n]{x^n+y^n}$ 
\item $+^s$ (shifted sum, $x+^c y:=x+y-1$, over whole $\R$, with neutral $1$)
\item $\times^c$ (complemented product, $x+y-xy$, can be extended also to whole $\R$, stabilizes the range of probabilities or fuzzy $[0,1]$ and 
is distributive over the shifted sum)
\end{enumerate}

\ss
A table $T$ is a two-rows array, the first row being filled with words taken in a given free monoid (see \cite{DHL}, \cite{LA} in this conference or 
\cite{Lo}). The set of words which are present in the first row will be called the {\it indices} of the table ($I(T)$) and for the second row the 
{\it values} or ({\it coefficients})of the table. The order of the columns is not relevant. Thus, a table reads

\begin{equation}
\left\{ 
\begin{array}{ll}
  indices & \mbox{set of words $I(T)$}\\
  values & \mbox{bottom row $V(T)$}
\end{array}\\
\right.
\end{equation}

The laws defined on tables will be of two types:\\ 
pointwise type (subscript $_p$) and convolution type (subscript $_c$).\\
Now, we can define the pointwise composition (or product) of two tables, noted $\ostar_p$.\\
Let us consider, two tables $T_1,\ T_2$ and a law $*$

\begin{center}
$T_1=$
\begin{tabular}{c|c|c|c}
$u_1$ 	& 	$u_2$ 	& $\cdots $ & $u_k$\\
\hline
$p_1$ 	& 	$p_2$ 	& $\cdots $ & $p_k$\\
\end{tabular}
\\
\vspace*{0.4cm}%\hs 
and
\vspace*{0.4cm}%\hs 
\\
$T_2=$
\begin{tabular}{c|c|c|c}
$v_1$ 	& 	$v_2$ 	& $\cdots $ & $v_l$\\
\hline
$q_1$ 	& 	$q_2$ 	& $\cdots $ & $q_l$\\
\end{tabular}
\end{center}

then $T_1\ostar_p T_2$ is defined by  $T_i[w]$ if $w\in I(T_i)$ and $w\notin I(T_{3-i})$
and by $T_1[w]* T_2[w]$ if $w\in I(T_1)\cap I(T_2)$

\ss
In particular one has $I(T_1\ostar_p T_2)=I(T_1)\cup I(T_2)$.

\begin{note}
i) At this stage one do no need any neutral. The structure automatically creates it (see {\rm algebraic remarks} below for full 
explanation).\\
ii) The above is a considerable generalization of an idea appearing in \cite{CD}, aimed only to semirings with units.
\end{note}

\ss
For convolution type, one needs two laws, say $\oplus, \otimes$, the second being distributive over the first, i.e. identically 
\begin{eqnarray}
x\otimes (y\oplus z)&=&(x\otimes y)\oplus (x\otimes z)\ \mathrm{and}\cr
(y\oplus z)\otimes x&=&(y\otimes x)\oplus (z\otimes x) 
\end{eqnarray}

(see\\
{\small {\tt http://mathworld.wolfram.com/\\
Semiring.html}}).\\
 
The set of indices of $T_1\ostar_c T_2$ ($I(T_1\ostar_c T_2)$) is the concatenation of the two (finite) langages $I(T_1)$ and $I(T_2)$ i.e. the (finite) 
set of words 
\begin{equation}\label{convolution}
I(T_1)I(T_2)=\{uv\}_{(u,v)\in I(T_1)\times I(T_2)}. 
\end{equation}

then, for $w\in I(T_1)I(T_2)$, one defines

\begin{equation}
T_1\otimes_c T_2[w]=\bigoplus_{uv=w}\Big( T_1[u] \otimes T_2[v]\Big)
\end{equation}

the interesting fact is that the constructed structure (call it $\T$ for tables) is then a semiring $(\T,\oplus_p,\otimes_c)$ 
(provided $\oplus$ is commutative and - generally - without units, but this is sufficient to perform matrix computations). There is, in fact no 
mystery in the definition \mref{convolution} above, as every table can be decomposed in elementary bits

\begin{equation}
T_1=
\begin{tabular}{c|c|c|c}
$u_1$ 	& 	$u_2$ 	& $\cdots $ & $u_k$\\
\hline
$p_1$ 	& 	$p_2$ 	& $\cdots $ & $p_k$\\
\end{tabular}
=\ \bigoplus_{i=1}^k
\begin{tabular}{|c|}
$u_i$\\
\hline
$p_i$\\
\end{tabular}
\end{equation}

one has, thanks to distributivity, to understand the convolution of these indecomposable elements, which is, this time, 
very natural

\begin{equation}
\begin{tabular}{|c|}
$u_1$\\
\hline
$p_1$\\
\end{tabular}
\ \bigotimes_c\ 
\begin{tabular}{|c|}
$u_2$\\
\hline
$p_2$\\
\end{tabular}
\ :=\ 
\begin{tabular}{|c|}
$u_1u_2$\\
\hline
$p_1\times p_2$\\
\end{tabular}
\end{equation}

\subsection{Why semirings ?}

In many applications, we have to compute the weights of paths in 
some weighted graph (shortest path problem, enumeration of paths, 
cost computations, automata, transducers to cite only a few) and the computation goes with two main rules: multiplication in series 
(i.e. along a path), and addition in parallel (if several paths are involved).\\
This paragraph is devoted to showing that, under these conditions, the axioms of Semirings are by no means arbitrary and in fact unavoidable.
A weighted graph is an oriented graph together with a {\it weight} mapping $\omega: A\mapsto K$ from the set of the arrows ($A$) to some set of 
coefficients $K$, an arrow is drawn with its weight (cost) above as follows $a=q_1\stackrel{\al}{\ra} q_2$.\\
For such objects, one has the general conventions of graph theory.
\begin{itemize}
\item $t(a):=q_1$ ({\it tail}) 
\item $h(a):=q_2$ ({\it head})
\item $w(a):=\al$ ({\it weight}).
\end{itemize}
A {\it path} is a sequence of arrows $c=a_1a_2\cdots a_n$ such that 
$h(a_k)=t(a_{k+1})$ for $1\leq k\leq n-1$. The preceding functions are extended to paths by
 $t(c)=t(a_1),\ h(c)=h(a_n)$, $w(c)=w(a_1)w(a_2) \cdots w(a_n)$ (product in the set of coefficients).\\

For example with a path of length 3 and ($k=\N$), 
\begin{equation}
u=p\stackrel{2}{\ra} q \stackrel{3}{\ra} r \stackrel{5}{\ra} s
\end{equation}
one has $t(u)=p,\ h(u)=s,\ w(u)=30$.

\ss
As was stated above, the (total) weight of a set of paths with the same head and tail is the sum of the individual weights. 
For instance, with

\begin{equation}
 {\mathbf q1}\mhs {\stackrel{\al}{\ra}  \atop \stackrel{\be}{\ra}}\mhs {\mathbf q2}
\end{equation}

the weigth of this set of paths est $\al+\be$. From the rule that the weights multiply in series and add in parallel one can 
derive the necessity of the axioms of the semirings. The following diagrams shows how this works. 

\begin{center}
\begin{tabular}{|l|l|}
\hline
Diagram 	& Identity 	\\
\hline
\hline
$\ \ \ \stackrel{\al}{\ra}$ &\\
$p  \stackrel{\be}{\ra} q$ & $\al+(\be+\ga)=(\al+\be)+\ga$ \\
$\ \ \ \stackrel{\ga}{\ra}$ &\\
\hline
$p {\stackrel{\al}{\ra} \atop \stackrel{\be}{\ra}} q$ & $\al+\be=\be+\al$\\
\hline
$p\stackrel{\al}{\ra} q \stackrel{\be}{\ra} r \stackrel{\ga}{\ra} s$ &
$\al(\be\ga)=(\al\be)\ga$ \\
\hline
$p {\stackrel{\al}{\ra} \atop \stackrel{\be}{\ra}} q \stackrel{\ga}{\ra} r$ & $(\al+\be)\ga=\al\ga+\be\ga$ \\
\hline
$p\stackrel{\al}{\ra} q {\stackrel{\be}{\ra} \atop \stackrel{\ga}{\ra}} r $ & $\al(\be+\ga)=\al\be+\al\ga$\\
\hline
\end{tabular}
\end{center}
these identities are familiar and bear the following names:
\begin{center}
\begin{tabular}{|l|l|}
\hline
Line & Name\\
\hline\hline
I & Associativity of $+$\\
\hline
II & Commutativity of $+$\\
\hline
III & Associativity of $\times$\\
\hline
IV & Distributiveness (right) of $\times$ over $+$\\
\hline
V & Distributiveness (left) of $\times$ over $+$\\
\hline
\end{tabular}
\end{center}

\subsection{Total mass}

The total mass of a table is just the sum of the coefficients in the bottom row. One can check that 
\begin{eqnarray}
mass(T1\oplus T2)&=&mass(T1)+mass(T2);\cr
mass(T1\otimes T2)&=&mass(T1) \cdot  mass(T2)
\end{eqnarray}
this allows, if needed, stochastic conditions.

\subsection{Algebraic remarks}

We have confined in this paragraph some proofs of structural properties concerning the tables. The reader may skip this section 
with no serious harm.\\
First, we deal with structures with as little as possible requirements, i.e. {\it Magmas} and {\it Semirings}. For formal 
definitions, see

{\small{\tt http://\\
encyclopedia.thefreedictionary.com/\\
Magma\%20category}

\ss
{\tt http://mathworld.wolfram.com/\\
Semiring.html}}\\

\begin{proposition} (i) Let $(S,*)$ be a magma, $\Si$ an alphabet, and denote $T[S]$ the set of tables with indices in $\Si^*$ and values in $S$. 
Define $\ostar_p$ as in \mref{table}. Then\\
i) The law $\ostar$ is associative (resp. commutative) iff $*$ is. 
Moreover the magma $(\T[S],\ostar)$ always possesses a neutral, the empty table (i.e. with an empty set  of indices).\\
ii) If $(K,\oplus,\otimes)$ is a semiring, then $(\T_K,\oplus,\otimes)$ is a semiring.\\
\end{proposition}

\Proof (Sketch) Let $S_{(1)}$ the magma with unit built over $(S\cup \{e\})$ by adjunction of a unit. Then, to each table $T$, associate the (finite 
supported) function $f_T: \Sigma^*\mapsto S_{(1)}$ defined by
\begin{equation}
f_T(w) = 
\left\{ 
\begin{array}{lll} 
  T[w] & \mbox{if $w\in I(T)$}\\
   e & \mbox{otherwise} 
\end{array}
\right.
\end{equation}
then, check that $f_{T_1\ostar_p T_2}=f_{T_1}\ostar_1 f_{T_2}$ (where $\ostar_1$ is the standard law on $S_{(1)}^{(\Sigma^*)}$) and that the 
correspondence is a isomorphism. Use a similar technique for the point (ii) with $K_{0,1}$ the semiring with units constructed over $K$ and show 
that the correspondence is one-to-one and has $K_{0,1}\langle \Sigma \rangle$ as image.

\begin{note} 1) Replacing $\Sigma^*$ by a simple set, the (i) of proposition above can be extended without modification (see also $K$-subsets in 
\cite{Ei}).\\
2) If one replaces the elements of free monoid on the top row by elements of a semigroup $S$ and admits some colums with a top empty 
cell, we get the algebra of $S_{(1)}$.\\ 
3) Pointwise product can be considered as being constructed with respect to the (Hadamard) coproduct $c(w)=w\otimes w$ whereas convolution 
is w.r.t. the Cauchy coproduct
\begin{equation}
c(w)=\sum_{uv=w}u\otimes v
\end{equation}  
(see \cite{DFLL}).
\end{note}

\section{Applications}
\subsection{Specializations and images}

\begin{enumerate}
\item {\bf Multiplicities, Stochastic and Boolean}\pointir

Whatever the multiplicities, one gets the classical automata by emptying the alphabet (setting $\Sigma =\emptyset$). For stochastic, one can use the 
total mass to pin up outgoing conditions.

\item {\bf Memorized Semiring}\pointir

We explain here why the memorized semiring, devised at first to perform efficient computations on the shortest path problem with 
memory (of addresses) can be considered as an image of a "table semiring" (thus proving without computation the central property 
of \cite{Ka}).\\
Let $\T$ be here the table semiring with coefficients in $([0,+\infty],min,+)$. Then a table 
\begin{equation}
T=
\begin{tabular}{c|c|c|c|c}
$u_1$ 	& $\cdots$ &	$u_k$ 	& $\cdots $ & $u_n$\\
\hline
$l_1$ 	& $\cdots$ &	$l_k$ 	& $\cdots $ & $l_n$\\
\end{tabular}
\end{equation}
can be written so that $l_1=\cdots=l_k<l_{m}$ for $m>k$ (this amounts to say that the set where the minimum is reached is 
$\{u_1,u_2\cdots u_k\}$). Then, to such a table, one can associate $\phi(T):=[\{u_1,u_2\cdots u_k\},l_1]$ in the memorized semiring. 
It is easy to check that $\phi$ transports the laws and the neutrals and obtain the result.
\end{enumerate}

\subsection{Application to evolutive systems}
Tables are structured as semirings and are flexible enough to recover and amplify the structures of automata with multiplicities and 
transducers. 
%we can apply to automata with mutations (see \cite{B1,B2}).
They give operational tools for modelling agent behaviour for various simulations in the domain of distributed artificial intelligence \cite{B2}.
The outputs of automata with multiplicities or the values of tables allow to modelize in some cases agent actions or in other cases, probabilities on possible transitions between internal states of agents behaviour.
In all cases, the algebraic structures associated with automata outputs or tables values is very interesting to define automatic computations in respect with the evolution of agents behaviour during simulation.\\

One of ours aims is to compute dynamic multiagent systems formations which emerge from a simulation. 
The use of table operations delivers calculable automata aggregate formation. 
Thus, when table values are probabilities, we are able to obtain evolutions of these aggregations as adaptive systems do.\\

With the definition of adapted operators coming from genetic algorithms, we are able to represent evolutive behaviors of agents and so evolutive systems \cite{B1}.
Thus, tables and memorized semiring are promizing tools for this kind of implementation which leads to model complex systems in many domains.

\end{multicols}

\begin{thebibliography}{ABC}
\bibitem{B1} Bertelle C., Flouret M., Jay V., Olivier D., Ponty J.-L., {\it Genetic Algorithms on Automata with Multiplicities for Adaptative 
Agent Behaviour in Emergent Organisations}.
\bibitem{B2} Bertelle C., Flouret M., Jay V., Olivier D., Ponty J.-L., {\it Automata with Multiplicities as Behaviour Model in Multi-Agent 
Simulations} SCI 2001.
 \bibitem{CD} Champarnaud J.-M., Duchamp G., {\it Derivatives of rational expressions\\ and related theorems}, 
T.C.S. {\bf 313} 31 (2004).
\bibitem{DHL} Duchamp G., Hatem Hadj Kacem, \'Eric Laugerotte, {\it On the erasure  of  several letter-transitions}, 
JICCSE'04
\bibitem{DFLL} Duchamp G., Flouret M., Laugerotte E., Luque J-G., {\it Direct and dual laws for automata with 
multiplicites}, Theoret. Comput. Sci. 267 (2001) 105-120.
\bibitem{Ei} Eilenberg S., {\it Automata, languages and machines, Vol A}, Acad. Press (1974).
\bibitem{LA} Laugerotte E., Abbad H., {\it Symbolic computation on weighted automata}, JICCSE'04.
\bibitem{Lo} Lothaire M., {\it Combinatorics on words}, Cambridge University Press (new edition), 1997.
\bibitem{Ka} Khatatneh K., {\it Construction of a memorized semiring}, DEA ITA Memoir, University of Rouen (2003).
\end{thebibliography}
\end{document}